\documentclass[pra,aps,floatfix,showpacs,tightenlines,superscriptaddress,amsmath,amssymb,showkeys,10pt,nofootinbib]{revtex4-2}

\usepackage{slashed}
\usepackage{xcolor}
\usepackage{graphicx}

\usepackage{hyperref}

\usepackage{notes2bib}

\newcommand{\be}{\begin{equation}}\newcommand{\ee}{\end{equation}}
\newcommand{\bea}{\begin{eqnarray}}\newcommand{\eea}{\end{eqnarray}}
\newcommand{\brr}{\begin{array}}\newcommand{\err}{\end{array}}
\newcommand{\bit}{\begin{itemize}}\newcommand{\eit}{\end{itemize}}
\newcommand{\ben}{\begin{enumerate}}\newcommand{\een}{\end{enumerate}}

\newcommand{\bbm}{\begin{bmatrix}}\newcommand{\ebm}{\end{bmatrix}}
\newcommand{\ba}{\begin{array}}
\newcommand{\ea}{\end{array}}
\newcommand{\G}{\textbf}

\newtheorem{mydef}{Definition}
\newtheorem{Lemma}{Lemma}
\newcommand{\bd}{\begin{mydef}} \newcommand{\ed}{\end{mydef}}
\newcommand{\bthe}{\begin{theorem}} \newcommand{\ethe}{\end{theorem}}
\newcommand{\ble}{\begin{Lemma}} \newcommand{\ele}{\end{Lemma}}

\def\ha{\frac{1}{2}}

\def\lan{\langle}
\def\lf{\left}

\def\pa{\partial}\def\ran{\rangle}

\def\ri{\right}
\def\al{\alpha}

\def\1{{_{1}}}\def\2{{_{2}}}

\def\noHe0{:\;\!\!\;\!\!:H_e(0):\;\!\!\;\!\!:}
\def\noHm0{:\;\!\!\;\!\!:H_\mu(0):\;\!\!\;\!\!:}

\def\lan{\langle}
\def\lf{\left}

\def\pa{\partial}\def\ran{\rangle}

\def\ri{\right}

\def\al{\alpha}

\def\1{{_{1}}}\def\2{{_{2}}}

\begin{document}

\title{General properties of the response function in a class of solvable non-equilibrium models}

\author{Federico Corberi}
\email{fcorberi@unisa.it}
\affiliation{Dipartimento di Fisica, Universit\`a di Salerno, Via Giovanni Paolo II 132, 84084 Fisciano (SA), Italy}
\affiliation{INFN Sezione di Napoli, Gruppo collegato di Salerno, Italy}

\author{Luca Smaldone}
\email{lsmaldone@unisa.it}
\affiliation{Dipartimento di Fisica, Universit\`a di Salerno, Via Giovanni Paolo II 132, 84084 Fisciano (SA), Italy}
\affiliation{INFN Sezione di Napoli, Gruppo collegato di Salerno, Italy}

\begin{abstract}
We study the non-equilibrium response function $R_{ij}(t,t')$, namely the variation of the local magnetization $\langle S_i(t)\rangle$ on site $i$ at time $t$ as an effect of a perturbation applied at the earlier time $t'$ on site $j$, in a class of solvable spin models characterized by the vanishing of the so-called {\it asymmetry}.
This class encompasses both systems brought out of equilibrium by the variation of a thermodynamic control parameter, as after a temperature quench, or intrinsically out of equilibrium models with violation of detailed balance. The one-dimensional Ising model and the voter model (on an arbitrary graph) are prototypical examples of these two situations which are used here as guiding examples. Defining the fluctuation-dissipation ratio $X_{ij}(t,t')=\beta R_{ij}/(\partial G_{ij}/\partial t')$, where $G_{ij}(t,t')=\langle S_i(t)S_j(t')\rangle$ is the spin-spin correlation function and $\beta$ is a parameter regulating the strength of the perturbation (corresponding to the inverse temperature when detailed balance holds), we show that, in the quite general case of a kinetics obeying dynamical scaling, on equal sites this quantity has a universal form
$X_{ii}(t,t') = (t+t')/(2t)$, whereas $
\lim _{t\to \infty}X_{ij}(t,t')=1/2$ for any $ij$ couple. The specific case of voter models with long-range interactions is thoroughly discussed.
\end{abstract}

\maketitle
\maketitle

\section{Introduction}
\label{sec:intro}

While statistical mechanics provides a rather complete and detailed theory of equilibrium systems, understanding out of equilibrium states is still an open challenge. Depending on the system at hand, in some cases such states can be treated as weak perturbations around a given equilibrium, allowing sometimes to build some kind of perturbation approach. However, there are in nature a wealth of phenomena that cannot be considered near to equilibrium by any means. One example is provided by systems building an ordered state out of an initially disordered one, as in the case of phase-ordering~\cite{Bray94,PuriWad09,CRPHYS_2015__16_3_257_0} in quenched magnets or phase-separation in composite materials as, for instance, binary liquids or alloys. Glassy materials~\cite{Bouchaud1997OutOE,Corberi2010GrowingLS} and other disordered systems~\cite{CORBERI2015332} also fall into the same category. These systems remain out of equilibrium for very long times because the ergodic time diverges in the thermodynamic limit and may therefore become larger than any accessible experimental timescale. Moreover, although the kinetics becomes slower and slower as time elapses, a phenomenon dubbed as {\it aging}, the probability distribution in phase-space keeps changing at any time as it can be revealed by inspection of suitable observables as, for instance, two-time correlations. This {\it slowly relaxing} state 
can be viewed as gradual exploration of larger and larger portion of phase-space in a progressive effort to reinstate ergodicity, a scenario referred to as {\it weak ergodicity breaking}. The problem, therefore, cannot be handled in a statistical mechanical approach by restricting over a subset of phase space, as it would be appropriate for the study of metastability~\cite{10.1007/BFb0025611}.

A different, but related, class of systems are those 
kept away from equilibrium by a small external power input done by stationary non-conservative forces violating detailed balance, as in active matter or in sheared or stirred materials, and/or periodically time-dependent forces. 

In this context, a wealth of studies have been concerned with the behavior of two-time quantities and, in particular, on the relation between the response function $R(t,t')$, with $t\ge t'$, and the associated correlation function
$A(t,t')$, the fluctuation-dissipation relation. Here $R(t,t')$ is the so-called autoresponse, namely the local reaction of the system at position $\vec x$ at time $t$, due to a perturbation acting in the same place at the earlier time $t'$. Similarly, $A(t,t')$ is the autocorrelation function between a local observable measured at $\vec x$ at two different times $t'$ and $t$.   
In equilibrium, the \emph{fluctuation-dissipation theorem} (FDT)~\cite{Kubo1991} provides a linear relation between these two quantities involving the inverse thermodynamic temperature $\beta=1/k_BT$
\begin{equation}
\beta^{-1}R(t,t')=\frac{\partial}{\partial t'}A(t,t').
\label{eqReq}
\end{equation}
In a canonical setting, this guarantees that a thermometer immersed in the system under study displays a temperature equal to the one of the bath the sample is in contact with~\cite{PhysRevE.55.3898}.
Out of equilibrium, the FDT does not hold and the quantity 
\be
\beta _{eff}=\beta X^{-1}(t,t')=\frac{\frac{\partial A(t,t')}{\partial t'}}{R(t,t')}
\label{eqX}
\ee
has been interpreted, in some cases, as an effective inverse temperature~\cite{Corberi_2004}, different from that of the reservoir. For instance, in a class of aging systems as, e.g., the $p$-spin model of a glass \cite{PhysRevLett.71.173,doi:10.1080/01418639508238541}, the so called \emph{fluctuation-dissipation ratio} (FDR) $X(t,t')$ takes, for large times $t'$, only a couple of values, $X(t,t')=1$ for $t-t'\ll t'$ and $X(t,t')=X_\infty\neq 1$ for $t-t'\gg t'$, showing the presence, next to the bath temperature, of an additional effective (inverse) temperature $\beta _{eff}=\beta X^{-1}_\infty$.
In addition, the FDR has been also used as an indicator of the
out-of-equilibrium nature of the system, and has been related to the rate of energy dissipation~\cite{PhysRevLett.95.130602} or the entropy production rate~\cite{PhysRevLett.79.2168}.

Interestingly, besides evidencing the nature of the deviation from equilibrium, in slowly relaxing systems $X(t,t')$ -- a quantity compute in a far-from-equilibrium state -- encodes also information on the equilibrium state the system is approaching.
Indeed, a theorem~\cite{PhysRevLett.81.1758,Franz1999} relates the FDR to the overlap probability distribution of the target equilibrium.

In this paper we discuss the form of the FDR in a class of models where the so-called {\it asymmetry} (see Sec.~\ref{sec:generalities}, Eq.~(\ref{asymdef}) in particular) vanishes, a fact that makes an analytical approach possible. Brownian diffusion is perhaps the simplest example.
The problem is defined in terms of spin variables $S_i=\pm 1$ located on the vertices $i$ of a completely general graph.
For concreteness, after discussing general properties, we will consider the voter model and perform a quantitative analysis on regular $D$-dimensional lattices, considering interactions depending on the distance
$r$ between spins~\cite{corberi2023kinetics,corsmal2023ordering,corberi2024aging,corberi2024coarseningmetastabilitylongrangevoter}. The one-dimensional case with nearest neighbors (NN) interactions can be mapped to the kinetic Ising chain, whose FDR during phase-ordering was derived long ago in Refs~\cite{Lippiello2000,Godrèche_2000}. 
Besides that, we will consider here cases with long-range interactions between spins decaying as $r^{-\alpha}$, for any $\alpha$. In the scaling regime of the model, where the time evolution can be reparametrized in terms of a single growing length $L(t)$ (see Sec.~\ref{sec:generalities}, especially around Eq.~(\ref{scalG} for a precise definition) the FDR is a non-trivial universal function
of its arguments decaying from $X\equiv 1$ at equal times $t=t'$, to $X\equiv 1/2$ for large time separations $t-t'\to \infty$. We also discuss the space-dependent FDR obtained through a generalization of Eqs.~(\ref{eqReq},\ref{eqX}) where the response describes the reaction of the system at $\vec x$ due to a perturbation acting earlier on a different site $\vec x +\vec r$. This quantity has a more complicted behavior than its equal site counterpart but, similarly to it, it also converges to $X=1/2$ for large $t-t'$.

The paper is organized as follows: in Section \ref{sec:generalities} we present some generalities on fluctuation-dissipation theorem and we derive some results on FDR which will be tested in the voter model. In Section \ref{sec:votermodel} we review the voter model with long-range interactions in its general form and then we provide explicit calulations of both response function and FDR in both one- and two-dimensions. Finally in Section \ref{sec:conlcusions} we present conclusions and discussion. For reader's convenience we have devoted Appendix \ref{appA} to derive some auxiliary computations which were employed in the main text. 
%%%%%%%%%%%%%%%%%%%%%%%%%%%%%%%%%%%%%%%%%%%%%%%%%%%%%%%%%%%%%%%%%%%%
\section{Fluctuation-dissipation relations: generalities} \label{sec:generalities}

In this Section we review some generalities 
on linear response theory and fluctuation-dissipation relations in and out of equilibrium. We will use the paradigm of spin systems,
but most of the properties reviewed here are rather general. 

Let us consider a collection of Ising variables $S_i=\pm 1$ evolving according to a Markovian stochastic process in discrete time. In general, the linear response function $R^{AB}_{ij}(t,t')$ describes the modification of an observable $A_i$ at site $i$ and time $t$ due to a small perturbation $b$ coupling linearly 
with another variable $B_j$ at previous times 
$t'$. The simplest instance, in a spin system, is when $A_i\equiv S_i$, $B_j\equiv S_j$ and 
$b=h_j$ is a magnetic field. When a kick perturbation acting on a very short time $\Delta t$ is applied 
\begin{equation}
	h_j(t)=h\theta (t-t')\theta(t'+\Delta t-t)\delta _{i,j},
\end{equation} 
where $\theta$ is the Heaviside function, the \emph{impulsive response} is defined as~\cite{ChristopheChatelain_2003,ACrisanti_2003}
\begin{equation}
	R_{ij(t,t')}=\lim _{\Delta t \to 0} \frac{1}{\Delta t}\left . \frac{\delta \langle S_i(t)\rangle}{\delta h_j(t')}\right | _{h=0}.
\end{equation}
Here $\langle \dots \rangle$ indicates a non-equilibrium statistical average, namely taken over initial configurations and stochastic realizations of the evolution, and
\begin{equation}
	\delta \langle S_i(t)\rangle=\langle S_i(t)\rangle_h-\langle S_i(t)\rangle
\end{equation}
where $\langle \dots \rangle_h$ means an average over trajectories in the presence of the perturbation.

The response function can be alternatively written as~\cite{Lippiello2000,Corberi2005fd,Corberi2010fd}
\begin{equation}
\beta ^{-1}R_{ij}(t,t')=\frac{1}{2}\left [\frac{\pa}{\pa t'}G_{ij}(t,t')
-\langle S_i(t)B_j(t')\rangle\right ],
\label{eqR}
\end{equation}
where 
\begin{equation}
	G_{ij}(t,t')=\langle S_i(t)S_j(t')\rangle
\end{equation}
is the spin-spin correlation function, and
\begin{equation}
    B_j=-2S_j w(S_j),
\end{equation}
$w(S_i)$ being the transition rate for flipping a single spin $S_i$ in the absence of the perturbation. Eq.~(\ref{eqR}) holds true for any kind of unperturbed dynamic rules, also if detailed balance is violated, for any perturbation
modifying the unperturbed transition rates as
$w(S_i)\to w^h(S_i)$ with
\begin{equation}
	w^h(S_i)=w(S_i)(1-\beta h_j),
	\label{detbalh}
\end{equation}
where $\beta $ is a parameter regulating the effect
of the kick. If the unperturbed evolution obeys detailed balance, $\beta$ can be identified with the inverse thermodynamic temperature $\beta =(k_BT)^{-1}$, $k_B$ being the Boltzmann constant.
Eq.~(\ref{detbalh}) can be interpreted as the 
requirement that the perturbed Markov chain obeys detailed balance {\it at least} with respect to the effect of a perturbation adding an energy 
$-h_jS_j$ to the unperturbed system. Clearly, if the latter obeys detailed balance with respect to a certain Hamiltonian, Eq.~(\ref{detbalh}) amounts to the natural request that detailed balance is still valid after switching on the perturbation. 
Notice that on the r.h.s. of Eq.~(\ref{eqR}) there are only correlation functions computed in the unperturbed dynamics.
Therefore Eq.~(\ref{eqR}) can be qualified as a non-equilibrium generalization of the FDT. Indeed, it is simple to show~\cite{Lippiello2000,Corberi2005fd,Corberi2010fd} that it reduces to the usual equilibrium relation 
\begin{equation}
	\beta ^{-1}R_{ij}(t,t')=\frac{\pa}{\pa t'}G_{ij}(t,t')
	\label{fdt}
\end{equation} 
in equilibrium conditions. This will be further discussed below.  Let us also mention that 
Eq.~(\ref{eqR}) also holds, with a suitable definition of $B_j$, for systems of continuous variables described, e.g., by Langevin or Fokker-Planck equations~\cite{PhysRevE.78.041120}. 

It is useful to re-write Eq.~(\ref{eqR}) in the
alternative form
\begin{equation} \label{1voterresp1}
	\beta^{-1}R_{ij}(t,t')=\frac{1}{2}
	\left \{\frac{\pa G_{ij}(t,t')}{ \pa t'} - \frac{\pa G_{ij}(t,t')}{ \pa t}
	-\mathcal{A}_{ij}(t,t')\right \},
\end{equation}
where 
\begin{equation}
	\mathcal{A}_{ij}(t,t')=\langle S_i(t)B_j(t')\rangle-\frac{\pa G_{ij}(t,t')}{ \pa t}
 \label{asymdef}
\end{equation}
is the so called {\it asymmetry}. Since it can be shown~\cite{PhysRevE.78.041120} that $\frac{\pa G_{ij}(t,t')}{ \pa t} =\langle B_i(t)S_j(t')\rangle$, this quantity amounts to a sort of commutator, 
\begin{equation}
    \mathcal{A}_{ij}(t,t')=\langle S_i(t)B_j(t')\rangle-\langle B_i(t)S_j(t')\rangle \, ,
\end{equation}
whereby the name comes from. In stationary states with time-inversion invariance, $\mathcal{A}_{ij}$ vanishes and one recovers the equilibrium FDT, Eq.~(\ref{fdt}).
However, equilibrium is not a necessary condition 
for having $\mathcal{A}_{ij}\equiv 0$, for instance the same holds true in the aging stage of the $1D$ Ising model~\cite{Lippiello2000,Godrèche_2000,Corberi2005fd} and,
as we will see soon, in the voter model in arbitrary 
space dimension and with arbitrary spacial interactions.

The generalized FDT~(\ref{eqR}) can be also written in the form \cite{Lippiello2000}
\be
\beta R_{ij}(t,t') \ = \ X_{ij}(t,t') \, \frac{\pa G_{ij}(t,t')}{ \pa t'}  \, , 
\ee
where
\be
X_{ij}(t,t') \ = \ \ha \, \lf[1-\frac{\frac{\pa G_{ij}(t,t')}{ \pa t} }{\frac{\pa G_{ij}(t,t')}{ \pa t'} }-\frac{\mathcal{A}_{ij}(t,t')}{\frac{\pa G_{ij}(t,t')}{ \pa t'}}\ri] \, 
\label{defX1}
\ee
is the FDR. This quantity can, in some cases, be associated to a non-equilibrium effective temperature~\cite{Corberi_2004}. Clearly, in equilibrium one has $X_{ij}(t,t')\equiv 1$. In the following we will denote the equal-site FDR $X_{ii}(t,t')$ as 
$X(t,t')$ {\it tout court}. Similarly, also the autocorrelation function 
$G_{ii}(t,t')$ will be denoted as $A(t,t')$.

The considerations above are general. Now we 
restrict our discussion to systems where the asymmetry vanishes. 
The FDR reads, in this case,
\be
X_{ij}(t,t') \ = \ \ha \, \lf[1-\frac{\frac{\pa G_{ij}(t,t')}{ \pa t} }{\frac{\pa G_{ij}(t,t')}{ \pa t'} }\ri ] \,. 
\label{defX}
\ee
The first observation is that, in stationary states where $G_{ij}$ is a function of the
time difference $t-t'$ only, 
one has $X_{ij}(t,t')\equiv 1$, as in equilibrium.
In systems where detailed balance is violated, such states are non-equilibrium 
ones and occur, for instance in the voter model, as we will discuss. In addition, it can be easily
proved that $\lim _{t'\to t}X(t,t')=1$, even in cases in which detailed balance does not hold (see Appendix~\ref{appA}). 
This can be clearly 
seen in Figs.~\ref{fig_X_variAlfa},\ref{fig_X_2d_variAlfa}, where $X(t,t')$ is plotted for the $d=1$ and $d=2$ voter model with long-range interactions (see Sec.~\ref{sec:votermodel}), respectively. In this figures, following established practice, $X(t,t')$ is plotted against 
$A(t,t')$. Given that $A(t,t')$ is a function decreasing monotonously from $A(t,t)=1$ to zero (for $t \gg t'$), the region $t'\to t$
corresponds to the right part of the plot.

Instead, if $i\neq j$ the behavior of $X_{ij}(t,t')$ in the above equal times limit is radically different, since one has $\lim _{t\to t'}X_{i\neq j}(t,t')=0$. This is shown in 
Appendix ~\ref{appA}. The physical intuition is that flipping $S_i$ at time $t$ or 
$S_j$ at time $t'$ occurs with the same probability, hence $\frac{\partial G_{ij}}{\partial t}=\frac{\partial G_{ij}}{\partial t'}$ in Eq.~(\ref{defX}), as $t'\to t$. Alternatively, one arrives at the same conclusion upon arguing that a perturbation 
exerted somewhere else at $j$, cannot reach
$i$ in a vanishing time interval. As a concrete example, this can be observed for the long-range $d=1$ voter model in Fig.~\ref{fig_X_vari_r}.

Now we focus on aging systems. Assuming spatial homogeneity it is $G_{ij}(t,t')=G(r;t,t')$,
where $r$ is the distance between $i$ and $j$. Generally the late-time dynamics obeys a dynamical scaling symmetry~\cite{Bray94,PuriWad09,CRPHYS_2015__16_3_257_0,FurukawaJStatSocJpn,PhysRevB.40.2341,TJNewman_1990,PhysRevE.52.6082,Corberi_2012,Chamon_2011,PhysRevE.65.046136,PhysRevLett.83.5054,PhysRevE.59.213} expressed by the form
\begin{equation}
	G(r;t,t')=g\left (\frac{r}{L(t')};\frac{L(t)}{L(t')}\right ),
	\label{scalG}
\end{equation}
where $L(t)$ is a dynamical correlation length. From Eq.~(\ref{defX}) one has
\begin{equation}
	X(r;t,t') \ = \ \ha \, \lf[1+\frac{\frac{dg}{dz}}{y\frac{dg}{dy}+z\frac{dg}{dz}}\,\frac{\frac{\pa L(t)}{\pa t}}{\frac{\pa L(t')}{\pa t'}}\ri] \,, 
	\label{scalX1}
\end{equation}
where $y=r/L(t')$ and $z=L(t)/L(t')$ are the arguments of $g$ in Eq.~(\ref{scalG}).
If, as it is usually the case, $L(t)$ grows algebraically in time, $L(t)\propto t^a$,
then it is
\begin{equation}
	X(r;t,t') \ = \ x(y;z) \ = \ \ha \, \lf[1+\frac{\frac{dg}{dz}}{y\frac{dg}{dy}+z\frac{dg}{dz}}\,z^{\frac{a-1}{a}}\ri] \, ,
	\label{scalX2}
\end{equation}
showing that $X$ is also a function $x$ of the scaling variables $y$ and $z$ alone.
Focusing on the case $r=0$, which is the situation most frequently considered, the above expression gives
\be \label{xtt'}
X(t,t') \ = \ \frac{1+z^{-\frac{1}{a}}}{2}=\frac{t+t'}{2 t}\, .
\ee
Then $X(t,t')$ is a universal quantity, independent on the specific model considered
among those with vanishing asymmetry, in a regime characterized by dynamical scaling. Let us just stress that, when plotting 
against $A(t,t')$ as it is done in Figs.~\ref{fig_X_variAlfa},\ref{fig_X_2d_variAlfa}, curves inherit the model dependence of $A(t,t')$, as it can be visually observed.  

On the contrary, for $r\neq 0$, Eq.~(\ref{scalX2}) shows that the model-dependent scaling function $g$ informs $X$. 
The limiting value~\cite{Godrèche_2000}
\begin{equation}
	X_\infty(r;t')=\lim_{t\to \infty}X(r;t,t'),
\end{equation}
(taken with fixed $r$ and $t'$), however, retains a universal character similar to
the one pointed out for $X(t,t')$.
In fact, the second term in square brackets in Eq.~(\ref{scalX2}) generally vanishes as $z\to \infty$ (because $z\frac{dg}{dz}\to 0$ and $a\le 1$), and hence
\begin{equation}
	X_\infty(r;t')\equiv \frac{1}{2},
\end{equation}
for any $r$ and $t'$. This is clearly seen in Fig.~\ref{fig_X_vari_r}.

%%%%%%%%%%%%%%%%%%%%%%%%%%%%%%%%%%%%%%%%%%%%%%%%%%%%%%
\section{The voter model} \label{sec:votermodel}
%%%%%%%%%%%%%%%%%%%%%%%%%%%%%%%%%%%%%%%%%%%

The voter model~\cite{Kimura1964,1970mathematics,Clifford1973,Holley1975,Clifford1973,Holley1975,liggett2004interacting,Theodore1986,Scheucher1988,PhysRevA.45.1067,Frachebourg1996,Ben1996,PhysRevLett.94.178701,PhysRevE.77.041121,Castellano09} is described by a set of binary variables located on the nodes $i$ of a graph, which can assume the values $S_i=\pm 1$. Two spins $S_i$ and $S_k$ interact with a probability $P_{i k}$, so that the probability to flip $S_i$ is
\be
w(S_i)=\frac{1}{2} \, \sum _k P_{i k} (1-S_iS_k) \, .
\label{transvoter}
\ee

With standard techniques of stochastic calculus~\cite{Glauber} it is easy to show (see Appendix~\ref{appA}) that 
$\frac{\pa}{\pa t}\langle S_{i}(t)S_{j}(t')\rangle=-2\langle S_i(t)S_j(t')w(S_{i}(t))\rangle$
which, provides
\begin{eqnarray}
\frac{\pa G_{ij} (t,t')}{\pa t}&=&-G_{ij}(t,t')+\sum _k P_{i k} G_{kj}(t,t') \, .
\label{eqc0}
\end{eqnarray}

Regarding the response function, using the transition rate $w$ of Eq.~(\ref{transvoter}),
Eq.~(\ref{eqR}) amounts to
\begin{eqnarray}
    \beta^{-1}R_{ij}(t,t')&=&
\frac{1}{2}\left [\frac{\pa G_{i j}(t,t')}{\pa t'}
+G_{ij}(t,t')-\sum _kP_{j k}
G_{ik}(t,t')\right ].
\end{eqnarray}
Comparing with Eq.~(\ref{eqc0}), and using the fact that $G_{ij}$ and $R_{ij}$ are symmetric in $i,j$,
it follows
\begin{equation} \label{1voterresp2}
    \beta^{-1}R_{ij}(t,t')=\frac{1}{2}
    \left \{\frac{\pa G_{ij}(t,t')}{ \pa t'} - \frac{\pa G_{ij}(t,t')}{ \pa t}
\right \} \, ,
\end{equation}
showing that the asymmetry is vanishing in this model.

%%%%%%%%%%%%%%%%%%%%%%%%%%%%%%%%%%%%%%%%%%%
\subsection{Voter model on a 
regular lattice} \label{sec:voterregular}
%%%%%%%%%%%%%%%%%%%%%%%%%%%%%%%%%%%%%%%%
We now focus on the case where the graph is a $D$-dimensional lattice and the probability $P_{ik}$ that spin $S_i$ interacts with $S_k$ depends only on their distance.
Considering the correlation 
$G_{ij}(t,t')=G(r;t,t')$, where 
$r$ is the distance between $i$ and $j$, and indicating with $d_k$ the distance between $k$ and $j$, the evolution equation for this quantity reads~\cite{corberi2023kinetics,corsmal2023ordering,corberi2024aging,corberi2024ordering,corberi2024coarseningmetastabilitylongrangevoter}
\begin{equation}
\frac{\pa G (r;t,t')}{\pa t} =-G (r;t,t') +\sum _\ell P(\ell) \sum _{k=1}^{n(\ell)}G (d_k(r,\ell),t)\, ,
\label{eqc1}
\end{equation}
where $\ell$ runs over all the possible distances on the lattice and
$n(\ell)$ is the number of lattice sites $k$ at distance $\ell$ from $i$. 
Eq.~(\ref{eqc1}) must be solved
with {\it initial} condition
$G(r;t,t)=C(r;t)$, where the 
equal-time correlation $C$ obeys~\cite{corberi2023kinetics,corsmal2023ordering,corberi2024aging} a differential equation analogous to Eq.~(\ref{eqc1}). We will consider the case of the ordering process originating from a fully disordered state at $t=0$.
Further details on this problem can be found in~\cite{corberi2024aging}.

In the following we will focus on the case of interactions decaying algebraically with distance, i.e.
\begin{equation}
P(\ell)\propto \ell^{-\alpha},
\label{eqP}
\end{equation}
with $\alpha >0$.
In the two opposite limits $\alpha \to \infty$ and $\alpha \to 0$ one recovers, respectively, the original voter model with NN interactions only, and the mean
field version where the agents are fully connected.

We start by considering the one-dimensional case where Eq.~(\ref{eqc1}) simplifies to
\begin{equation}
\frac{\pa G (r;t,t')}{\pa t} =-G (r;t,t') +\sum _{\ell=1}^{N/2} P(\ell)\,[G(r-\ell;t,t')+G(r+\ell;t,t')]\, .
\label{eqc11}
\end{equation}
In the NN limit $\alpha \to \infty$ the model can be mapped exactly onto the Ising model. 
Therefore, in this case (but only in this case) the model obeys detailed balance. More in general, the above equation
was studied with periodic boundary conditions in~\cite{corberi2024aging}, where different behaviors were found as $\alpha $ is changed. In the following, 
building on this knowledge, we discuss the behavior of the response function. We will always assume the thermodynamic limit $N\to \infty$.

\begin{figure}[h]
	\vspace{1.0cm}
	\centering
	\rotatebox{0}{\resizebox{0.5\textwidth}{!}{\includegraphics{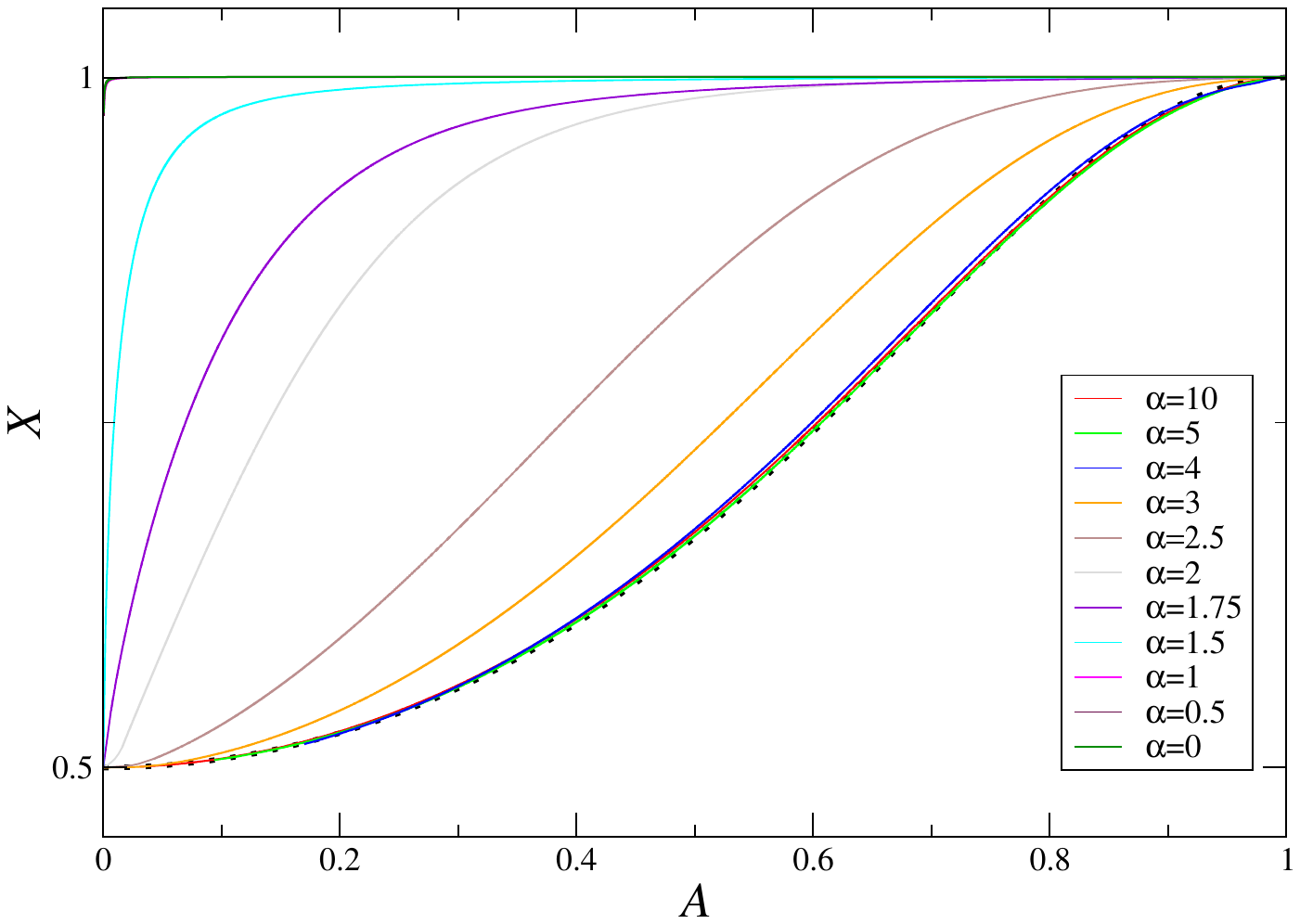}}} 
  \caption{$X(t,t')$ is plotted against $G(t,t')$, for the $1D$ voter model with long-range algebraic interactions. different curves are for different values of $\al$, indicated in the legend, for $t'=10$ ($t'=10^2$ for $\al=3$). System size is $N=10^3$ for $\al \ge 3$, $N=10^4$ for $2\le \al <3$, $N=10^5$ for $\al <2$. The black heavy-dotted curve is the analytical form for the 1d-Ising model with NN interactions obtained in~\cite{Lippiello2000,Godrèche_2000}. Curves for $\al =0,0.5,1$ superimpose.}
	\label{fig_X_variAlfa}
\end{figure}

For $\al >3$, $G(r;t,t')$ has the form \eqref{scalG}, with
$L(t)\propto t^{1/2}$, for sufficiently small values $y<y^*$ of $y=r/L(t')$ (this will be further detailed later on).
Then Eq.~(\ref{xtt'}) holds true.
The autocorrelation function $A(z)$ reads~\cite{corberi2024aging}
\be
A(z) \ = \ \frac{2}{\pi} \,  \arctan \left(\frac{\sqrt{2}}{\sqrt{z^2-1}}\right) \, , 
\label{LZ}
\ee
which is the same result as in the NN model \cite{Lippiello2000,Godrèche_2000} and is independent on $\al$. 
Inverting Eq.~(\ref{LZ}) to obtain $z(A)$ and plugging 
into Eq.~(\ref{defX}) one has~\cite{Lippiello2000,Godrèche_2000}
\be
X(A) \ = \ \frac{1}{2-\sin^2 \lf(\frac{\pi}{2} A\ri)}\,, \hspace{2cm} \alpha >3\, . 
\label{XA}
\ee
This analytic determination is plotted with a black heavy-dotted curve in Fig.~\ref{fig_X_variAlfa}. 
In the same figure, continuous curves are obtained by numerically solving Eq.~(\ref{eqc11}), thus obtaining $A(t,t')$ by letting $r=0$ in $G(r;t,t')$ and in $X(r;t,t')$, the latter quantity being obtained, starting from $G(r;t,t')$, by means of Eq.~(\ref{defX}). One sees that the curves for $\alpha >3$ are $\alpha$-independent and fall onto the analytic form~(\ref{XA}).

\begin{figure}[h]
	\vspace{1.0cm}
	\centering
	\rotatebox{0}{\resizebox{0.5\textwidth}{!}{\includegraphics{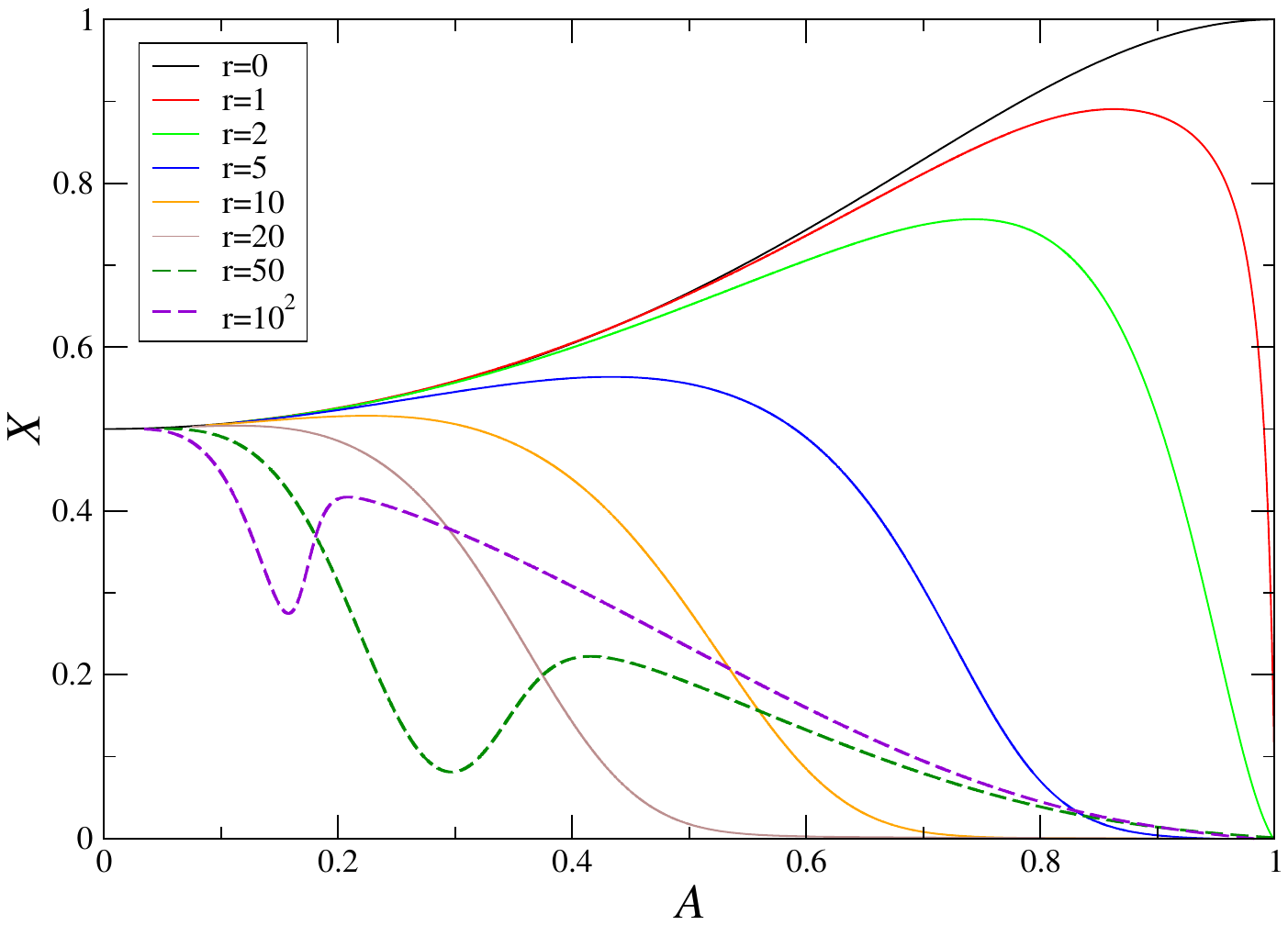}}} 
  \caption{$X(r;t,t')$ is plotted against $A(t,t')$, for $\alpha =5$ and different values of $r$, indicated in the legend. 
  System size is $N=10^3$ and
  $t'=10$.}
	\label{fig_X_vari_r}
\end{figure}

We now move to study $X(r;t,t')$ for finite values of $r$. To the best of our knowledge this quantity has not been thoroughly studied before. Its behavior, for $\alpha =5$, is shown in Fig.~\ref{fig_X_vari_r}, where it is plotted against $A(t,t')$, as we already did for $X(t,t')$ in Fig.~\ref{fig_X_variAlfa}. 
As we mentioned in Sec.~\ref{sec:generalities}, $X(r;t,t')$ 
is bound to take the values $X(r;t,t)=0$ for equal times (right part of the figure, $A= 1$), and $X(r;t\to \infty,t')=1/2$ in the large time separation sector (left part of the plot, $A=0$). This is clearly observed in Fig.~\ref{fig_X_vari_r}.  
It can also be noticed that the curves approach the one for $r=0$ (black line) 
below a characteristic value $A^*$.
Looking at Eq.~(\ref{scalX2}), this means that the term $z\,dg/dz$ in the denominator on the r.h.s. eventually prevails over 
$y\,dg/dy$. Clearly, this occurs at values of $z$ (i.e. of $t$) larger as $y$ (hence $r$) are taken larger, as indeed we see in Fig.~\ref{fig_X_vari_r}. Moreover, we notice a rather different form of such approach for small ($r<r^*$, with $ r^* \simeq 25$) of large ($r>r^*$) values of $r$ (to better display this we draw data for $r<r^*$ with continuous lines  
and those for $r>r^*$ with dashed ones).
For $r<r^*$ the curves rise to a maximum and then approach the curve for $X(t,t')$
(i.e. black) smoothly from above. For $r>r^*$, instead, there is an additional wiggle. The origin of $r^*$ and its impact on the form of $X(r;t,t')$ can be understood as follows. 
How we anticipated earlier, a unique scaling form as in Eq.~(\ref{scalG}) is obeyed by $G(r;t,t')$ only for sufficiently small values $y<y^*$ of $y=r/L(t')$. Indeed, for larger values of
$y$, both the growth law of the scaling length and the form of the scaling function $g$ change significantly~\cite{corberi2024aging}.
In particular, while for $y<y^*$ both $g$ and $L(t)$ behave as in the NN case, for $y>y^*$ they become $\al$-dependent. 
The value $y^*$ separating these different
behaviors is an increasing function $y^*(t',z)$ of $z$ and $t'$ and diverges as $\alpha \to \infty$, meaning that for NN interactions a unique scaling form is recovered.  
Plotting $X(r;t,t')$ for fixed $t'$ and a small value of $r$, such that $r<y^*(t',z=1)=r^*$, the condition $r<y^*(t',z)$
will remain true for all the subsequent evolution, meaning that Eq.~(\ref{defX})
only probes the $y<y^*(t',z)$ behavior of 
$G(r;t,t')$. This produces the (continuous) lines for $r<r^*$ shown in Fig.~\ref{fig_X_vari_r}. On the other hand, considering $X(r;t,t')$ for a value of $r$ as large as to have $r>y^*(t',z=1)=r^*$, Eq.~(\ref{defX}) initially probes the large-$y$ $\alpha$-dependent scaling of $G(r;t,t')$.
As time elapses (and $z$ increases), since $y^*(t',z)$ is an increasing function of $z$, one crosses over to the universal scaling form valid for $y<y^*(t',z)$. This produces the wiggle observed in Fig.~\ref{fig_X_vari_r}. Clearly, the larger is the value of $r$ considered, the later the wiggle will be produced, which is indeed observed in Fig.~\ref{fig_X_vari_r}. It is interesting 
to notice the effectiveness of $X(r;t,t')$ in providing such detailed information on the spatial structure of correlations in the systems. Indeed, the existence of $y^*$ can be hardly inferred by the direct study of correlation functions, particularly if numerical, because correlations are already very small at $y^*$ and masked by any source of noise. Instead, $X(r;t,t')$ displays a very clear (additional) non-monotonicity. Let us also comment that such an effect is a peculiar feature of the extended interactions present in the model, since there is no such a feature in the NN case (because $y^*=\infty$).

The quantitative discussion above holds true for $\alpha >3$. Using the results contained in~\cite{corberi2024aging}, and proceeding similarly, one can extend the previous analysis to the range with $\alpha \le 3$. Specifically, for $2< \alpha \le 3$ and large $t'$ and $z$, the autocorrelation function reads
\be
A(z) \ = \ a_1 \, z^{-1}  \, ,
\label{factscal}
\ee
where $a_1$ is a constant, with
$L(t)\propto t^{1/(\alpha-1)}$.
Then one has
\be
X(A) \ = \ \frac{1+(A/a_1)^{\al-1}}{2}  \, . 
\label{X23}
\ee
This equation shows that $X(A)$ decreases
significantly towards the asymptotic value 
$X(0)=1/2$ when $(A/a_1)^{\alpha -1}$ gets smaller than a certain number $\epsilon \ll 1$. Solving we get $A\ll a_1\epsilon ^{1/(\alpha -1)}$, thus implying that the
descent occurs at smaller and smaller values of $A$ as $\alpha$ decreases, which is indeed seen in Fig.~\ref{fig_X_variAlfa}. 

In the range $1< \alpha \le 2$ the system undergoes a coarsening phase with
\be
A(z) \ = \ a_2 \, z^{-\frac{1}{\al-1}} \, ,
\label{factscala2}
\ee
with another constant $a_2$ and
with
$
L(t) \ \propto \ t,
$
while approaching to a non-trivial stationary state, characterized by
\be
A(t,t')=A_{stat}(\tau) \ = \ A_0 \  \tau^{\frac{\al-2}{\al-1} } \, , \quad \tau \equiv t-t' \, .
\ee
Then, one has
\be
X(A) \ = \ \frac{1+(A/a_2)^{\al-1}}{2}  \, . 
\label{X12}
\ee
during such coarsening stage. Notice that 
the same form holds (apart from the value of the constants $a_1,a_2$) both for 
$2<\alpha \le 3$ (Eq.~(\ref{X23})) and for 
$1<\alpha \le 2$ (Eq.~(\ref{X12})) despite the fact that both the growth law $L(t)$ and the autocorrelation behave differently.
As a consequence, the curves for $X(t,t')$
in Fig.~\ref{fig_X_variAlfa} behave similarly to those of the cases with $2<\alpha \le 3$ and keep staying closer
to $X=1$ down to smaller and smaller $A$-values as $\alpha$ is progressively lowered, for the reasons previously discussed.

Later, when the system enters the stationary state, one has
\be
R(t,t')=R_{stat}(\tau) \ = \  \beta \, A_0 \,  \frac{ (\alpha -2) \tau^{\frac{1}{1-\alpha }}}{\alpha -1} \, , 
\ee
and $ X(r;t,t') \ \equiv \ 1$ trivially, as in any stationary state.

For $\al \le 1$, the situation is different because the stationary state is approached in a microscopic time, without any previous coarsening stage~\cite{corberi2023kinetics}.
At stationarity the system behaves similarly to the mean-field model, with
\be
A(t,t')=A_{stat}(\tau) \ = \ a_3 \, e^{-\tau}  \, ,
\ee
$a_3$ being a constant.
Then
\be
R_{stat}(\tau) \ = \  \beta \, A(\tau)\, , 
\ee
and $ X(r;t,t') \ \equiv \ 1$.
This can be seen in Fig.~\ref{fig_X_variAlfa}.

\begin{figure}[h]
	\vspace{1.0cm}
	\centering
	\rotatebox{0}{\resizebox{0.5\textwidth}{!}{\includegraphics{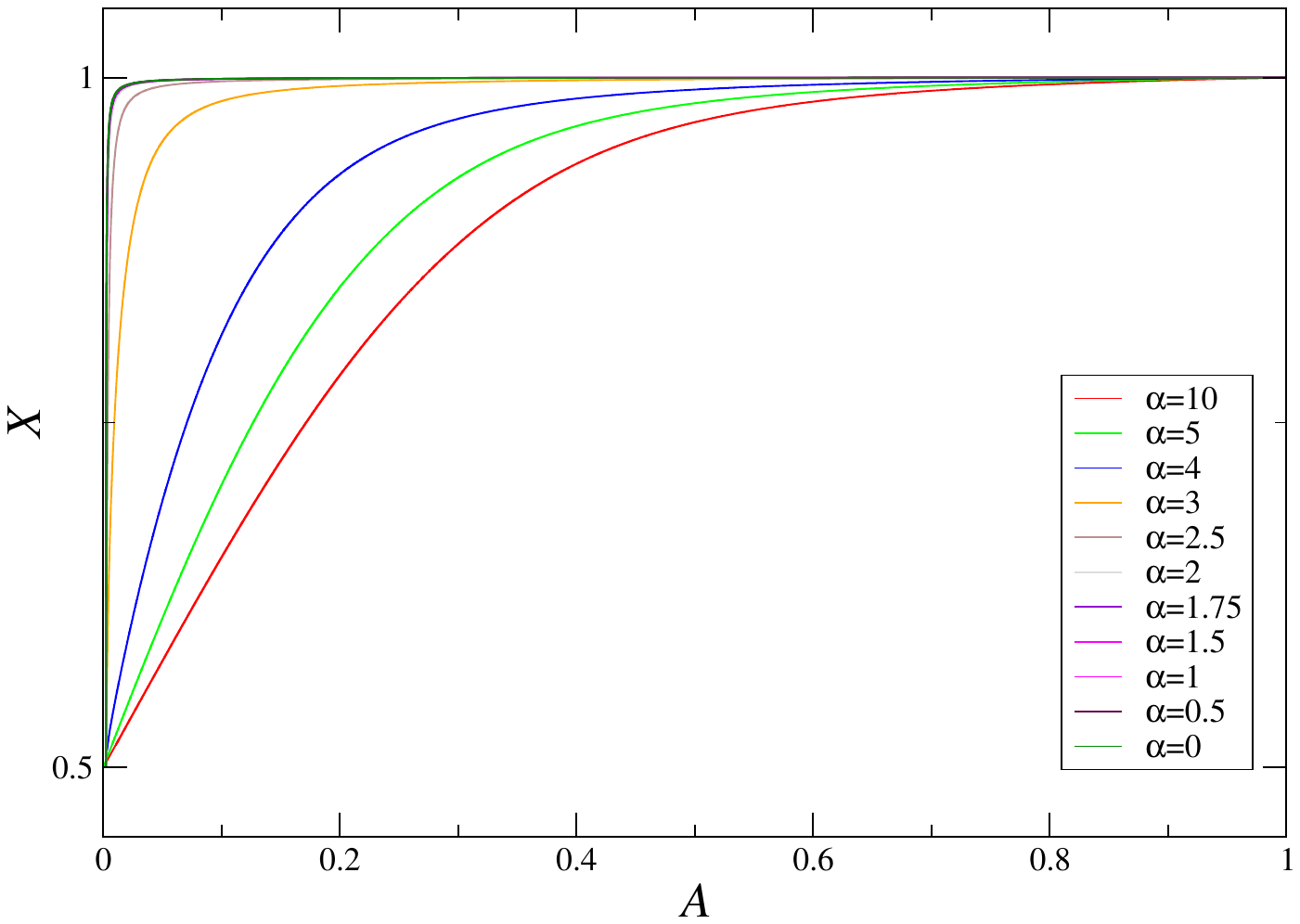}}} 
  \caption{$X(t,t')$ is plotted against $A(t,t')$, for the $2D$ voter model with long-range algebraic interactions. Different curves are for different values of $\al$, indicated in the legend, for $t'=10$. System size is $N=101^2$. Curves for $\alpha \le 2$ superimpose.}
	\label{fig_X_2d_variAlfa}
\end{figure}

We complete our discussion of the voter model on a lattice with interaction probability decaying algebraically as in Eq.~(\ref{eqP}) by briefly considering the 
case of a two-dimensional square lattice.
This model has been studied in~\cite{corsmal2023ordering} but two-time quantities, in particular $G(r;t,t')$ have never been analytically determined.
Our numerical results, obtained by numerically solving Eq.~(\ref{eqc1}) with periodic boundary conditions, are displayed in Fig.~\ref{fig_X_2d_variAlfa}.  The overall behavior is similar to the one-dimensional case, with $X(t,t')$ decaying from $X(t,t)=1$ to $X(t\to \infty,t')=1/2$ on smaller and smaller values of $A$ as $\alpha $ is decreased. Also, the mean-field behavior with $X(t,t')\equiv 1$ is observed for 
$\alpha \le 2$ since, in any dimension, this is expected for $\alpha \le D$. At variance with the $D=1$ case, however, $X$ depends on $\alpha$ in the whole range explored and a short-range universal behavior where $X$ is $\alpha$ independent, corresponding to $\alpha >3$ for $D=1$, is not observed.

%%%%%%%%%%%%%%%%%%%%%%%%%%%%%%%%%%%%%%
\section{Conclusions} \label{sec:conlcusions}

In this work, we studied the impulsive response function $R_{ij}(t,t')$ for a class of spin systems characterized by vanishing asymmetry and subjected to a small magnetic perturbation. The voter and the Ising model can be viewed as prototype examples. We also derived some general properties of the FDR $X_{ij}(t,t')$ for this class of models. Specifically, for systems exhibiting dynamical scaling in their kinetics, we found $X_{ii}(t,t') = (t+t')/(2t)$ and  $
\lim _{t\to \infty}X_{ij}(t,t')=1/2$ for any $ij$ pair. Building on the results from Ref. \cite{corberi2024aging}, which provided the analytical form of two-time autocorrelation functions for the one-dimensional voter model with long-range interactions, we calculated the response function and the FDR using the FDT in its general form~(\ref{eqR}), for spin systems. Moreover, we provided numerical results for both one and two-dimensional voter model.

The behavior of the response function and the associated FDR
has been a widely studied in aging system in the last years~\cite{Lippiello2000,Godrèche_2000,Corberi_2004,ChristopheChatelain_2003,ACrisanti_2003,Corberi2005fd,Corberi2010fd,Corberi_2012,Chamon_2011,PhysRevE.65.046136,PhysRevE.74.041106,PhysRevE.70.017103,PhysRevLett.90.099601,PhysRevE.72.056103,CorberiCugliandolo2009,Lippiello2006TestOL,90c20e4e-3345-36fa-a898-8d0bf35bb5ef,Franz1995,EnzoMarinari_1998,EnzoMarinari_2000,PhysRevB.67.214425}. One important direction in these studies is the possibility to infer static properties of hard-to-equilibrate systems, as glasses and spin glasses, from their dynamical response, enforcing the static-dynamic connection~\cite{PhysRevLett.81.1758,Franz1999}
mentioned in Sec.~\ref{sec:intro} which, under certain hypotheses, relates the out-of-equilibrium FDR to the equilibrium structure of phase space. However, it was found that in the one-dimensional Ising model with NN interactions one of those hypotheses is violated. Indeed, the FDR in that case (discussed in Sec.~\ref{sec:voterregular}) has an higly non-trivial form, completely independent of the trivial properties of the equilibrium low-temperature states. The origin of such violation is in the nature of the domain walls motion in $D=1$, which is purely diffusive, since there is no surface tension on the line. This produces an anomalously large response~\cite{PhysRevE.65.046114}, thus invalidating the static-dynamic connection.
Such connection, however, is restored when considering the same Ising model in $D>1$, because, in that case, surface tension regularizes the motion of the interfaces. The voter model considered in this paper is special because surface tension is absent in any dimension and, for this reason, it displays a non-trivial FDR qualitatively similar to the one of the NN Ising model, in any dimension (see Sec.~\ref{sec:voterregular}, Figs.~\ref{fig_X_variAlfa},\ref{fig_X_2d_variAlfa}, in particular). This is presumably true for the whole class of models with vanishing asymmetry considered in this paper.

%%%%%%%%%%%%%%%%%%%%%%%%%%%%%%%%%%%%%%%%%%%%%%%%%
\section*{Acknowledgments}

F.C. acknowledges financial support by MUR PRIN 2022 PNRR.

%%%%%%%%%%%%%%%%%%%%%%%%%%%%%%%%%%%%%%%%%%%%%%%%%%%%
\appendix
\section{Time derivatives of correlation functions} \label{appA}
In this Appendix we evaluate some time derivatives of $G_{ij}(t,t')$ for a general 
system of discrete variables (e.g. spins)
denoted by $\G S \equiv \lf(S_1, \ldots \, S_N\ri)$, 
subjected to a Markov process.
By definition
\be
G_{i j}(t,t') \ = \ \sum_{\G S,\G S'} \, S_i S'_j \, p\lf(\G S,t|\G S',t'\ri) \, p\lf(\G S',t'\ri)
\ee
where $p(\G S,t)$ is the probability of state $\G S $ at time $t$ and $p\lf(\G S,t|\G S',t'\ri) $ is the conditional probability of having the state $\G S$ at time $t$ if we had the state $\G S'$ at time $t'$. 

The increment of $G_{ij}$ when the largest time $t$ goes from $t$ to $t+\delta$, is
\be
\Delta_tG_{ij}(t,t')=\langle S_i(t+\delta)S_j(t')\rangle -\langle S_i(t)S_j(t')\rangle,
\ee
which explicitly reads
\be
\Delta_{t} G_{i j}(t,t') \ = \ \sum_{\G S,\G S',\G S''} \, (S''_i-S_i) S'_j \, p\lf(\G S'',t+\delta|\G S,t\ri) \, p\lf(\G S,t|\G S',t'\ri) \,  p\lf(\G S',t'\ri) \, .
\label{deltatG}
\ee
In the limit of small $\delta$ a single spin-flip attempt is only possible. Moreover, it is clear that only flip of $S_i$ contributes to the r.h.s. of Eq.~(\ref{deltatG}), due to the $S''_i-S_i$ factor. Therefore 
we can make the replacement $p(\G S'',t+\delta|\G S,t)=w(S_i) \delta$, leading to
\be
\frac{\pa G_{i j}(t,t')}{\pa t} \ = \ \lim _{\delta \to 0}\frac{\Delta_t G_{ij}(t,t')}{\delta} = \  \sum_{\G S, \G S',\G S''} \, (S''_i-S_i) \, S'_j  \,w\lf(S_i \ri) \, p\lf(\G S,t| \G S',t'\ri) \,  p\lf(\G S',t'\ri) \ \, .
\label{eqA5int}
\ee
For Ising spins ($S_i=\pm 1$) the only contribution is for $S''_i=-S_i$ hence, performing the sum over $\G S''$ one arrives at 
\be
\frac{\pa G_{i j}(t,t')}{\pa t} \ = \ -2 \, \sum_{\G S,\G S'} \, S_i \, S'_j  \,w\lf(S_i \ri) \, p\lf(\G S,t| \G S',t'\ri) \,  p\lf(\G S',t'\ri) \ = \ -2 \, \lan w(S_i(t)) S_i(t) S_j(t')\ran\, .
\label{eqA5}
\ee
This shows that the derivative of a correlator with respect of the largest time is a (different) correlator itself.

Let us now consider the limit $t'\to t$, where we can make the substitution 
$p\lf(\G S,t|\G S',t'\ri)\to \delta_{\G S, \G S'} $ and use this $\delta$-function to perform the $\G S'$ summation. One has
\be
\lim_{t'\to t}\frac{\pa G_{i j}(t,t')}{\pa t} \ = \ -2 \, \sum_{\G S} \, S_i \, S_j  \,w\lf(S_i \ri) \, p\lf(\G S,t\ri)  \ = \ -2 \, \lan  S_i(t) S_j(t)w(S_i(t))\ran\, .
\label{eqA5a}
\ee
In particular, one has
\be
\lim _{t'\to t} \frac{\pa G_{ii}(t,t')}{\pa t} \ =
\ -2\langle w(S_i(t))\rangle,
\ee
which simply expresses that for each spin flip, which occurs with probability $w(S_i)$, the correlation 
decreases by two.

Let us now consider the increment with respect to the smallest time $t'$, which is given by
\be
\Delta_{t'} G_{i j}(t,t') \ = \ \sum_{\G S,\G S',\G S''} \, S_i \lf(S''_j-S'_j\ri) \, p\lf(\G S,t|\G S'',t'+\delta \ri) \, p\lf(\G S'',t'+\delta|\G S',t'\ri) \,  p\lf(\G S',t'\ri).
\ee
Taking the incremental ratio $\frac{\Delta G_{i j}(t,t')}{\delta}$, proceeding as before and performing the limit $\delta\to 0$ we get
\be
\frac{\pa G_{i j}(t,t')}{\pa t'} \ = \ -2 \, \sum_{\G S,\G S'} \, S_i \, S'_j \, \,w\lf(S'_j \ri)p\lf(\G S,t| F_j \G S',t'\ri)  \,  p\lf(\G S',t'\ri) \, ,
\label{eqA3}
\ee
where we have defined the operator $F_j$ which flips the $j$-th spin, i.e. $F_j\G S\equiv (S_1,S_2,\dots,-S_j,\dots,S_N)$.
 This expression shows that the derivative of a correlator with respect to the smallest time cannot in general be written as a correlator, because of the $F_j$ in the conditional probability.
 
Let us now consider the limit $t'\to t$, where we can make the substitution 
$p\lf(\G S,t| F_j \G S',t'\ri)\to \delta_{\G S, F_j\G S'} $.
Using this $\delta$ function to perform the $\G S'$ summation in Eq.~(\ref{eqA3}) one has
\be
\lim_{t'\to t}\frac{\pa G_{i i}(t,t')}{\pa t'} \ = \left \{ \begin{array}{ll}
\ 2 \, \sum_{\G S} \,w\lf(S_i \ri) \,  p(\G S,t) \,=2\langle w(S_i(t)\rangle, \quad \quad & \mbox{for } i=j,\\
\ -2 \, \sum_{\G S} \, S_i \, S_j  \,w\lf(S_i \ri) \, p\lf(\G S,t\ri)  \ = \ -2 \, \lan  S_i(t) S_j(t)w(S_i(t))\ran, \quad \quad & \mbox{for } i\neq j,
\end{array}
\right .
\label{eqA3ttii}
\ee
showing that, only  
for equal times, also this derivative can be 
expressed in the form of an average quantity.
Comparing with Eq.~(\ref{eqA5a}) one has
\be
\lim_{t\to t'}\frac{\pa G_{ij}(t,t')}{\pa t'} \ = \left \{ \begin{array}{ll}
\ - \lim_{t\to t'}\frac{\pa G_{i i}(t,t')}{\pa t}, \quad \quad &\mbox{for } i=j, \\
\  \lim_{t\to t'}\frac{\pa G_{ij}(t,t')}{\pa t}, \quad \quad &\mbox{for } i\neq j.
\end{array} \right .
\ee
Notice that this result is independent on 
the detailed balance holding true or not and
applies to intrinsically non-equilibrium 
models as well. Plugging into Eq.~(\ref{defX}) this shows that $\lim _{t'\to t}X(t,t')=1$, while $\lim _{t'\to t}X_{ij}(t,t')=0, \forall i\neq j$.

%%%%%%%%%%%%%%%%%%%%%%%%%%%%%%%%%%%%%%%%%%%%%%%%%%%%%%%%%%
\section*{References}

\bibliography{LibraryStat}

\bibliographystyle{apsrev4-2}

\end{document}